\documentclass[showpacs,preprintnumbers,amsmath,amssymb,aps,prd,
               nofootinbib,eqsecnum,a4paper]{revtex4}
\usepackage[T2A]{fontenc}

\usepackage[cp1251]{inputenc}

\begin{document}
\title{To the study of non-Gaussianity in two-field slow-roll inflation.}

\author{N.A. Koshelev}
\email{koshna71@inbox.ru}

\affiliation{Ulyanovsk State University, Leo Tolstoy str 42,
432970, Russia}

\date{\today}

\begin{abstract}

The general expression for the second order large scale curvature
perturbation in the form of a functional over a background
solution is derived. The explicit expressions was obtained for two
special forms of the inflationary potential. In the considered
cases, it is shown that a significant level of non-Gaussianity can
be generated during the super-Hubble evolution only if
nonadiabatic perturbations are non-negligible at the end of
inflation.

\end{abstract}


\maketitle

\section{Introduction.}
\label{intro}

Primordial non-Gaussianity has emerged as one of important probes
of the inflationary era. In single field slow-roll inflation, the
non-Gaussianity is suppressed by the slow-roll parameters
\cite{Maldacena}, \cite{Seery_Lidsey1}, \cite{Seery_Lidsey2}, but
it can be sizeable in models with a transient violation of
standard slow-roll (for example,  due to a step feature in the
inflationary potential \cite{11103050}). Non-adiabatic
perturbations produced during multi-field inflation may lead to
generation of detectable deviations from Gaussian distribution
after inflationary stage. For instance, it can take place at an
inhomogeneous end of inflation \cite{0207295}, \cite{0510443},
\cite{12107818}, at the reheating \cite{0306006}, \cite{0303614},
\cite{0303591} or in the curvaton scenario \cite{0208055}.

It has also shown that significant non-Gaussianity can be obtained
during slow-roll multi-field inflation \cite{Alabidi}, \cite{BCH}.
An important feature of most of the known examples of this type is
that large scale curvature perturbation evolves even at the end of
the inflationary stage (see, however, \cite{11062153}). Hence,
such models are incomplete without an understanding of the
evolution of the cosmological perturbations until they become
nearly adiabatic \cite{10114934}. In particular, it is necessary
to consider the non-Gaussianity behavior at reheating. This raises
the question about principal possibility of slow-roll models with
large non-Gaussianity and adiabatic spectrum.

Most computations of the non-Gaussianity in multi-field inflation
have been carried out in the framework of the $\delta
\mathcal{N}$-formalism \cite{Starobinsky}, \cite{Sasaki_Stewart},
\cite{Sasaki_Tanaka}, \cite{0411220}, \cite{0504045}, using
labeling the trajectories by slow-roll integrals of motion
\cite{9511029}, \cite{Vernizzi_Wands}, \cite{11045238}. In an
attempt to go beyond standard assumptions, some extensions of the
$\delta \mathcal{N}$-formalism are worked out \cite{0506262},
\cite{10041870}, \cite{12106525}. Also, the covariant formalism
\cite{0610064}, \cite{08102405}, \cite{10033270}, \cite{Watanabe}
and the long-wavelength formalism \cite{0504508}, \cite{0506704},
\cite{0511041}, \cite{10126027} are developed. Although all
approaches mentioned above are geometrically transparent, here we
use method of employing the slow-roll Klein-Gordon equations.

The structure this paper is as follows. Chapter \ref{basic}
describes the model and basic equations. In Chapter \ref{basicSL},
slow-roll equations are  written and general slow-roll expression
for the curvature perturbation in two-field inflation is obtained.
For cases of product and sum potentials, the corresponding
expressions are written explicitly. In Chapter \ref{fnlcec}, using
decomposition of perturbations on the adiabatic and entropy ones,
it is shown that large values of $ f_{NL} $ can be generated only
at sufficiently large entropy perturbations  at the {\it{end}} of
inflationary stage. We summary our results in Section \ref{concl}.

\section{Model and basic equations.}
\label{basic}

Let us consider two canonical scalar fields $\varphi$ and $\chi$
minimally coupled to gravity
\begin{equation}
S = \int \left\{ -\frac{R}{16\pi G} - \frac{1}{2}\varphi ^{;\mu}
\varphi _{;\mu} - \frac{1}{2}\chi ^{;\mu} \chi _{;\mu} - U(
\varphi, \chi) \right\}\sqrt { - g} d^{4}x .
\end{equation}

The Lagrangian density is given by
\begin{equation}
\label{action} \mathcal {L} =- \frac{1}{2}\varphi ^{;\mu} \varphi
_{;\mu} - \frac{1}{2}\chi ^{;\mu} \chi _{;\mu} - U( \varphi, \chi)
.
\end{equation}

The Hilbert  energy-momentum tensor is defined as
\begin{equation}
\label{em_def}  T^{\mu\nu}=-2\frac{\partial \mathcal {L}}{\partial
g_{\mu\nu}} +g^{\mu\nu}\mathcal {L}  ,
\end{equation}
which gives
\begin{equation}
\label{em_fields}T^{\mu}_{~\nu}=\varphi^{;\mu}\varphi_{;\nu}
+\chi^{;\mu}\chi_{;\nu} -\delta^{\mu}_{~\nu}\left( U( \varphi,
\chi) + \frac{1}{2}\varphi^{;\mu}\varphi_{;\mu} +
\frac{1}{2}\chi^{;\mu}\chi_{;\mu} \right).
\end{equation}

The energy-momentum tensor (\ref{em_fields}), at least up to
second order perturbations, can also be treated as energy-momentum
tensor of a perfect fluid
\begin{equation}
\label{em_fluid}T^\mu_{~ ~\nu}= \left(\rho+P\right)u^\mu u_\nu + P
\delta^{\mu}_{~\nu} ,
\end{equation}
where $ \rho $ is density, $ P $ is pressure and $ u^\mu $ is
4-speed.

We split any quantity $f$ into a homogeneous background and small
inhomogeneous perturbations
\begin{equation}
f(\eta , x^i)=f^{(0)}(\eta)+\delta f^{(1)}(\eta , x^i) +
\frac{1}{2}\delta f^{(2)}(\eta , x^i)+ ... ~.
\end{equation}
where $\eta$ is conformal time, the Latin index $i$ takes values
from 1 to 3, and indices in brackets indicate the order of
perturbations.

Including second order perturbations, the line element around a
spatially flat Friedmann-Robertson-Walker background has the form
\cite{MW}
\begin{equation}
ds^2 =-a^2 (\eta)\left\{(1+2\phi^{(1)} +\phi^{(2)})d\eta^2 +
(2B^{(1)}_{i}+  B^{(2)}_{i})d\eta d x^i +\left[(1 - 2\psi^{(1)}
-\psi^{(2)}) \delta _{ij} + 2E^{(1)}_{ij}+
E^{(2)}_{ij}\right]x^ix^j\right\}.
\end{equation}

The metrics perturbations can be classified into scalar, vector,
and tensor  types  according to their transformation properties
under spatial coordinate transformations on constant-time
hypersurface \cite{Bardeen}. Here we will consider only the scalar
type perturbations, i.e., we set
\begin{equation}
B^{(1)}_{i}=B^{(1)}_{,i}, ~~~~~~B^{(2)}_{i}=B^{(2)}_{,i}, ~~~~~~
E^{(1)}_{ij}=E^{(1)}_{,ij}, ~~~~~~E^{(2)}_{ij} =E^{(2)} _{,ij}.
\end{equation}

The freedom of coordinate choice can be used to impose gauge
constraints . For example, one can use the uniform density gauge
($ \delta \rho^{(1)} = \delta \rho^{(2)} = 0 $) or the uniform
curvature gauge ($ \psi^{(1)} = \psi^{(2)} = 0 $).

In the analysis of perturbations, a special role is played by the
quantities that conserved on large scales if the pressure
perturbation is adiabatic. The first conserved nonlinear
gauge-invariant quantity was obtained in ref. \cite{MW}. The
commonly used quantity is curvature perturbation $ \zeta $ on
uniform density hypersurfaces, which is ambiguously defined in
nonlinear case. We adopt the definition of refs. \cite{Maldacena},
\cite{0502578}, \cite{0509078}. On large scales, this quantity is
associated with the variable of Malik and Wands \cite{MW}
\begin{equation}
\zeta^{(1)}_{MW} = -\left.\psi^{(1)}\right|_{\rho_{(1)} =
\rho_{(2)} = 0},~~~~~\zeta^{(2)}_{MW} =
-\left.\psi^{(2)}\right|_{\rho_{(1)} = \rho_{(2)} = 0}
\end{equation}
by relations \cite{0502578}, \cite{0509078}
\begin{equation}
\zeta^{(1)} = \zeta^{(1)}_{MW},~~~~~ \zeta^{(2)} =\zeta^{(2)}_{MW}
- {\zeta^{(1)}_{MW}}^2.
\end{equation}

Gauge transformations allow to write the second order curvature
perturbation $ \zeta^{(2)} $ in the gauge-invariant form
\cite{0502578}
\begin{equation}
\label{zeta2} \zeta^{(2)} = -\psi^{(2)}- \frac{ \mathcal{H}
}{\rho_{(0)}'}\delta\rho^{(2)} +2\frac{ \mathcal{H}}
{(\rho_{(0)}')^2}\delta\rho^{(1) \prime}\delta\rho^{(1)}
+2\frac{\delta\rho^{(1)}} {\rho_{(0)}'} \psi^{(1)\prime}- \left(
\mathcal{H} \frac{\rho_{(0)}''}{\rho_{(0)}'} -
\mathcal{H}'\right)\left(
\frac{\delta\rho^{(1)}}{\rho_{(0)}'}\right)^2.
\end{equation}
where $ \mathcal{H}=a'/a $ and prime denotes the derivative with
respect to conformal time $\eta$.  Most simply, this expression
appears in the uniform curvature ($UC$) gauge
\begin{equation}
\label{zeta2UC} \zeta^{(2)} = -\frac{ \mathcal{H} }{\rho_{(0)}'}
\delta\rho_{UC}^{(2)} + 2\frac{ \mathcal{H}} {(\rho_{(0)}')^2}
\delta \rho_{UC}^{(1) \prime}\delta\rho_{UC}^{(1)} - \left(
\mathcal{H} \frac{ \rho_{(0)}''}{\rho_{(0)}'} -
\mathcal{H}'\right)\left( \frac{ \delta
\rho_{UC}^{(1)}}{\rho_{(0)}'}\right)^2.
\end{equation}
In the following, all calculations done in the uniform curvature
gauge with conditions $ E^{(1)} = E^{(2)} = 0 $, what fix a
residual freedom of gauge transformations. In order to avoid
cluttering the expressions with too many subscripts, we omit the
index $"UC"$ below.

\subsection{Density perturbations.}

Let us write only a few components of the two-field
energy-momentum tensor in the uniform curvature gauge. Without
imposing gauge conditions, all components of the energy-momentum
tensor up to second-order terms can be found in Ref.
\cite{MW_rev}.

The perturbation expansion of  energy-momentum tensor
(\ref{em_fields}) yield
\begin{eqnarray}
\label{em000fields}T^{0}_{(0)0}&=& -\frac{1}{2a^2}{\varphi_{(0)}
'}^2 - \frac{1}{2a^2}{\chi_{(0)} '}^2 - U_{(0)} \\
\label{emij0fields}T^{i}_{(0)j}&=& \left[\frac{1}{2a^2}{
\varphi_{(0)}'}^2 + \frac{1}{2a^2}{ \chi_{(0)} '}^2 - U_{(0)}
\right]\delta^i_{~j} \\
\label{em001fields}\delta
T^{0}_{(1)0}&=&\frac{1}{a^2}\left[\left({\varphi_{(0)} '}^2+ {\chi
_{(0)}'}^2\right)\phi_{(1)} - \varphi_{(0)}' \delta\varphi_{(1)}
'- \chi_{(0)}' \delta\chi_{(1)} ' - a^2 U_{,\varphi}
\delta\varphi_{(1)} - a^2 U_{,\chi} \delta\chi_{(1)}\right] \\
\label{em002fields}\delta T^{0}_{(2)0} &=&
\frac{1}{a^2}\left[\left({\varphi_{(0)} '}^2+ {\chi_{(0)}
'}^2\right)\phi_{(2)} - \varphi_{(0)}' \delta\varphi_{(2)} ' -
\chi_{(0)}' \delta\chi_{(2)} '- a^2
U_{,\varphi}\delta\varphi_{(2)}- a^2 U_{,\chi}\delta\chi_{(2)}
-{\delta\varphi_{(1)}'} ^2 -
{\delta\chi_{(1)}'} ^2\right. \nonumber  \\
&-& \delta\varphi_{(1),i}\delta\varphi_{(1)}^{,i}
-\delta\chi_{(1),i}\delta\chi_{(1)}^{,i} + \left({\varphi_{(0)}
'}^2+ {\chi _{(0)}'}^2\right) B_{(1),i}B_{(1)}^{,i} +4 \left(
\varphi_{(0)} '\delta\varphi_{(1)}' + \chi_{(0)} '
\delta\chi_{(1)}'\right) \phi_{(1)} \nonumber \\
&-& \left.  a^2U_{,\varphi\varphi} \delta \varphi_{(1)}^2-
a^2U_{,\chi\chi} \delta \chi_{(1)}^2 - 2a^2U_{,\varphi\chi} \delta
\varphi_{(1)}\delta\chi_{(1)}- 4\left({\varphi _{(0)}'}^2+ {\chi
_{(0)}'}^2\right)\phi_{(1)}^2 \right], \\
\label{em0i1fields}\delta T^{0}_{(1)i} &=&-\frac{1}{a^2} \left(
\varphi_{(0)}' \delta\varphi_{(1),i} +\chi_{(0)}' \delta
\chi_{(1),i} \right),
\end{eqnarray}
where  we use the shorthand $U_{,\varphi}\equiv\frac{\partial
U_{(0)} }{\partial\varphi_{(0)}}$, $U_{,\chi}\equiv\frac{\partial
U_{(0)} }{\partial\chi_{(0)}}$, $U_{,\varphi\varphi} \equiv
\frac{\partial^2 U_{(0)}}{\partial\varphi_{(0)}\partial
\varphi_{(0)} }$, $U_{,\chi\chi} \equiv \frac{\partial^2
U_{(0)}}{\partial\chi_{(0)}\partial \chi_{(0)} }$,
$U_{,\varphi\chi} \equiv \frac{\partial^2
U_{(0)}}{\partial\varphi_{(0)}\partial \chi_{(0)} }$.

The alternative notation (\ref{em_fluid}) leads to expressions
\begin{eqnarray}
\label{em000fluid}T^{0}_{(0)0}&=&  - \rho_{(0)}, \\
\label{emij0fluid}T^{i}_{(0)j}&=&  P_{(0)}\delta^i_{~j}, \\
\label{em001fluid}\delta T^{0}_{(1)0} &=& -\delta \rho_{(1)} ,\\
\label{em002fluid}\delta T^0_{(2)0} &=& -\delta \rho_{(2)}
-2\left( \rho_{(0)}+P_{(0)}\right) v_{(1),i}v^{,i}_{(1)}
-2\left(\rho_{(0)}+P_{(0)}\right) B_{(1),i} v^{,i}_{(1)}, \\
\label{em0i1fluid}\delta T^{0}_{(1)i} &=& (\rho_{(0)}+P_{(0)})
\left(B_{(1),i} +v_{(1),i} \right),
\end{eqnarray}
where $ v $ is the velocity potential.

The comparison of equations (\ref{em000fields}),
(\ref{emij0fields}), (\ref{em0i1fields}) and (\ref{em000fluid}),
(\ref{emij0fluid}), (\ref{em0i1fluid}) gives
\begin{eqnarray}
\rho_{(0)} &=& \frac{1}{2a^2}{\varphi_{(0)}
'}^2 + \frac{1}{2a^2}{\chi_{(0)} '}^2 + U_{(0)}, \\
P_{(0)} &=& \frac{1}{2a^2}{\varphi_{(0)}
'}^2 + \frac{1}{2a^2}{\chi_{(0)} '}^2 - U_{(0)}, \\
v_{(1)}  &=& -B_{(1)} -  \frac{\varphi _{(0)}' \delta\varphi_{(1)}
+\chi _{(0)}' \delta\chi_{(1)}}{{\varphi _{(0)}'}^2+ {\chi_{(0)}
'}^2} .
\end{eqnarray}

From equations (\ref{em001fields}), (\ref{em002fields}) and
(\ref{em001fluid}), (\ref{em002fluid}), it follows now that
\begin{eqnarray}
\delta \rho_{(1)} &=& \frac{1}{a^2}\left[-\left({\varphi_{(0)}
'}^2+ {\chi_{(0)} '}^2\right)\phi_{(1)} + \varphi_{(0)}'
\delta\varphi_{(1)} '+ \chi_{(0)}' \delta\chi_{(1)} ' + a^2
U_{,\varphi} \delta\varphi_{(1)} + a^2 U_{,\chi} \delta\chi_{(1)}
\right], \\
\delta \rho _{(2)}   &=& \frac{1}{a^2}\Big[\varphi_{(0)}'
\delta\varphi_{(2)} ' + \chi_{(0)}' \delta\chi_{(2)} '+ a^2
U_{,\varphi}\delta\varphi_{(2)}+ a^2 U_{,\chi}\delta\chi_{(2)}
-\left({\varphi_{(0)} '}^2+ {\chi_{(0)} '}^2\right)
\phi_{(2)}  \nonumber \\
&-&\left({\varphi_{(0)} '}^2+ {\chi_{(0)} '}^2\right) \left(
B_{(1),i}B_{(1)}^{,i} - 4\phi_{(1)}^2 \right)+
{\delta\varphi_{(1)}'} ^2 + {\delta\chi_{(1)}'} ^2 -4
\left(\varphi _{(0)}' \delta\varphi_{(1)}' + \chi _{(0)}'
\delta\chi_{(1)}'\right) \phi_{(1)}\nonumber \\
&+& a^2U_{,\varphi\varphi} \delta \varphi_{(1)}^2+
a^2U_{,\chi\chi} \delta \chi_{(1)}^2+ 2a^2U_{,\varphi\chi} \delta
\varphi_{(1)}\delta\chi_{(1)}  - 2B_{(1)}^{,i}
\left(\varphi_{(0)}' \delta \varphi_{(1),i} +
\chi_{(0)}' \delta \chi_{(1),i}\right)  \nonumber \\
&-& 4 \frac{ \varphi_{(0)}' \chi_{(0)}' \delta \varphi_{(1)}^{,i}
\delta \chi_{(1),i} }{{ \varphi_{(0)} '}^2 +  {\chi_{(0)} '}^2} -
\frac{ {\varphi_{(0)} '}^2  - {\chi_{(0)}'}^2 }{{\varphi
_{(0)}'}^2 + {\chi _{(0)}'}^2} \left( \delta\varphi_{(1)
,i}\delta\varphi_{(1)}^{,i} - \delta \chi_{(1),i}
\delta\chi_{(1)}^{,i}\right) \Big].
\end{eqnarray}

\subsection{Perturbed Einstein equations.}

At first order, the $0-0$ component of the Einstein equations can
be written as
\begin{equation}
\label{first00}2a^2U_{(0)}\phi_{(1)}+\varphi'\delta\varphi_{(1)}'
+\chi'\delta\chi_{1}'+ a^2 U_{,\varphi}\delta\varphi_{(1)}+ a^2
U_{,\chi}\delta\chi_{(1)} +\frac{\mathcal{H}}{4\pi G}\Delta
B_{(1)} =0,
\end{equation}
where $\Delta $ is the Laplace operator.

The $0-i$ component is
\begin{equation}
\label{first0i}\mathcal{H}\phi_{(1)}=4\pi
G\varphi_{(0)}'\delta\varphi_{(1)} + 4\pi
G\chi_{(0)}'\delta\chi_{(1)}.
\end{equation}

The consideration of the off-diagonal $i-j$ components gives
\begin{equation}
\label{firstijoff}B_{(1)}'+2\mathcal{H}B_{(1)} +\phi_{(1)}=0.
\end{equation}
From the trace of $i-j$  components we find
\begin{eqnarray}
\mathcal{H}\phi_{(1)}' &+& (2\mathcal{H}'+\mathcal{H}^2)\phi_{(1)}
+\frac{1}{2} \left(\Delta\phi_{(1)}+2\mathcal{H}\Delta
B_{(1)}+\Delta B_{(1)}'\right)\nonumber\\
\label{firstijtr} &=& 4\pi G\left(- a^2 U_{,\varphi}
\delta\varphi_{(1)} - a^2 U_{,\chi} \delta\chi_{(1)} +
\varphi_{(0)}' \delta\varphi_{(1)} ' + \chi_{(0)}'
\delta\chi_{(1)} ' -\left({\varphi _{(0)}'}^2+ {\chi
_{(0)}'}^2\right)\phi_{(1)}\right).
\end{eqnarray}
Using equation (\ref{firstijoff}) and  background Einstein
equations, the latter can be reduced to
\begin{equation}
\label{firstijtr2}\mathcal{H}\phi_{(1)}' =4\pi G\left(- a^2
U_{,\varphi} \delta\varphi_{(1)} - a^2 U_{,\chi} \delta\chi_{(1)}
+ \varphi_{(0)}' \delta\varphi_{(1)} ' + \chi_{(0)}'
\delta\chi_{(1)} ' - 2U_{(0)} \phi_{(1)}\right).
\end{equation}

At second order, the $0-i$ Einstein equation is given by
\begin{equation}
\mathcal{H}\phi_{(2),i}-4\mathcal{H}\phi_{(1)}\phi_{(1),i}+2\mathcal{H}
B_{(1),ki}B_{(1)}^{~,k} +B_{(1),ki}\phi_{(1)}^{~,k}-\Delta
B_1\phi_{1,i}\nonumber
\end{equation}
\begin{equation}
\label{second0i}=4\pi G
\left[\varphi_{(0)}'\delta\varphi_{2,i}+\chi_{(0)}'\delta\chi_{(2),i}
+2\delta\varphi_{(1)}'\delta\varphi_{(1),i}+2\delta\chi_{(1)}'
\delta\chi_{(1),i}\right].
\end{equation}
We then use the first order $0-i$ equation (\ref{first0i}) and
take the trace of (\ref{second0i}), which gives \cite{Malik}
\begin{eqnarray}
\label{second0imod}\phi_{(2)} + B_{(1),k}B_{(1)}^{~,k} &=&
2\phi_{(1)}^2 +\frac{4\pi G}{\mathcal{H}} \left(\varphi_{(0)}'
\delta\varphi_{(2)} + \chi _{(0)}' \delta\chi_{(2)}\right)
-\frac{1}{\mathcal{H}}\Delta ^{-1} \left(\phi_{(1),kl} B_{(1)}
^{~~,kl} -\Delta B_{(1)}\Delta\phi_{(1)} \right) \nonumber \\
 &+& \frac{8\pi G}{\mathcal{H}} \Delta^{-1}\left( \delta\varphi_{(1)}'
\Delta \delta\varphi_{(1)} + \delta\chi_{(1)}' \Delta \delta
\chi_{(1)} +\delta\varphi_{(1),k}' \delta \varphi_{(1)} ^{~~,k}
+\delta\chi_{(1), k}' \delta \chi_{(1)} ^{~~, k} \right),
\end{eqnarray}
where $\Delta^{-1}$ is the inverse Laplacian, $\Delta^{-1}(\Delta
f)=f$.

\section{Slow-roll.}
\label{basicSL}

As usual, we introduce the two-field slow-roll parameters
\begin{equation}
\label{slowpar}\epsilon_\varphi \equiv \frac{1}{16\pi G} \left(
\frac{U_{, \varphi}}{U_{(0)}} \right)^2 ,~~~\epsilon_\chi \equiv
\frac{1}{16\pi G} \left( \frac{U_{, \chi}}{U_{(0)}} \right)^2,~~~
\eta_{\varphi\varphi} \equiv \frac{1}{8\pi G} \frac{U_{, \varphi
\varphi}}{U_{(0)}} ,~~~\eta_{\varphi\chi} \equiv \frac{1}{8\pi G}
\frac{U_{, \varphi \chi}}{U_{(0)}},~~~\eta_{\chi\chi} \equiv
\frac{1}{8\pi G} \frac{U_{, \chi\chi}}{U_{(0)}}.
\end{equation}

The slow-roll condition
\begin{equation}
\max \{\epsilon_\varphi, \epsilon_\chi, |\eta_{\varphi\varphi} |,
|\eta_{\varphi\chi} |, |\eta_{\chi\chi} |\} \ll 1
\end{equation}
leads to the relations
\begin{equation}
\varphi_{(0)}''-\mathcal{H}\varphi_{(0)}'\simeq 0, \qquad
\chi_{(0)}''-\mathcal{H}\chi_{(0)}'\simeq 0, \qquad
\frac{1}{2a^2}{\varphi_{(0)}'}^2 +
\frac{1}{2a^2}{\chi_{(0)}'}^2\ll U_0 .
\end{equation}
On large scales, for non-decaying modes of perturbations, there
are analogues
\begin{equation}
\label{slow1}\delta\varphi_{(1)}''-\mathcal{H}\delta\varphi_{(1)}'\simeq
0, \qquad \delta\chi_{(1)}''-\mathcal{H}\delta\chi_{(1)}'\simeq 0.
\end{equation}
\begin{equation}
\label{slow2}\delta\varphi_{(2)}''-\mathcal{H}\delta\varphi_{(2)}'\simeq
0, \qquad \delta\chi_{(2)}''-\mathcal{H}\delta\chi_{(2)}'\simeq 0.
\end{equation}
The basic slow-roll background equations are
\begin{equation}
\label{slowbasic1}3\mathcal{H}\varphi_{(0)}'+a^2U_{,\varphi}=0,
\qquad 3\mathcal{H}\chi_{(0)}'+a^2U_{,\chi}=0,
\end{equation}
\begin{equation}
\label{slowbasic2}\mathcal{H}^2=\frac{8 \pi G}{3}a^2 U_{(0)}.
\end{equation}
Also, the following slow-roll expressions are useful
\begin{equation}
\rho_{(0)} = U_{(0)},~~~~~~~~~\rho_{(0)} '=  -\frac{a^2}{
3\mathcal{H}}\left(U_{,\varphi}^2 +U_{,\chi}^2 \right),
\end{equation}
\begin{equation}
\mathcal{H} \frac{\rho_{(0)}''}{\rho_{(0)}'} - \mathcal{H}'
=\frac{a^2}{3} \frac{U_\varphi^2 +U_\chi^2}{U_{(0)}} -
\frac{2a^2}{3} \frac{U_{,\varphi}^2 U_{,\varphi \varphi}
+2U_{,\varphi} U_{,\chi}U_{,\varphi\chi} +U_{,\chi}^2 U_{,\chi
\chi}}{ U_{,\varphi}^2 +U_{,\chi}^2 }~.
\end{equation}

\subsection{Perturbed slow-roll Klein-Gordon equations.}

At first order, for $ N $ scalar fields $ \phi_I $ with the
potential $U(\phi_I)$, $I=1,...,N$, the Klein-Gordon equations can
be written as \cite{TaruyaNambu}
\begin{equation}
\label{KGN1}\delta\phi_{I(1)}''+2\mathcal{H}\delta\phi_{I(1)}'-\Delta
\delta\phi_{I(1)} +a^2\sum_K\left\{ U_{,\phi_K\phi_I} -\frac{8 \pi
G}{a^2} \left( \frac{a^2\phi_{K(0)}'\phi_{I(0)}'}{\mathcal{H}}
\right)' \right\}\delta\phi_{K(1)}=0.
\end{equation}

Considering the non-decaying modes of perturbations on large
scales, these equations are reduced in the slow-roll to
\begin{equation}
\label{KGN1ls}3\mathcal{H}\delta\phi_{I(1)}' +\sum_K \left(a^2
U_{,\phi_I \phi_K} -{24 \pi G}\phi_{I(0)}'\phi_{K(0)}'\right)
\delta \phi_{K(1)}=0.
\end{equation}
For two scalar fields $\varphi$ and $\chi$, the equations
(\ref{slowbasic1}), (\ref{slowbasic2}) and (\ref{KGN1ls}) give

\begin{eqnarray}
\label{phi1slow}\delta\varphi_{(1)}' &=& \frac{a^2}{3\mathcal{H}}
\left(\frac{ U_{,\varphi} U_{,\varphi}}{U}- U_{,\varphi \varphi}
\right) \delta \varphi_{(1)} + \frac{a^2}{3\mathcal{H}}
\left(\frac{U_{,\varphi}U_{,\chi}}{U}- U_{,\varphi \chi} \right)
\delta \chi_{(1)} , \\
\label{chi1slow}\delta\chi_{(1)}' &=& \frac{a^2}{3\mathcal{H}}
\left( \frac{ U_{,\chi}U_{,\varphi}}{U}- U_{,\chi \varphi} \right)
\delta \varphi_{(1)} + \frac{a^2}{3\mathcal{H}}\left( \frac{
U_{,\chi}U_{,\chi}}{U}- U_{,\chi \chi} \right) \delta \chi_{(1)}.
\end{eqnarray}

At second order, it is convenient to go to the Fourier
representation
\begin{equation}
f(\eta,x^i) = \int d^3 k \delta f_{\mathbf{k}} \exp\left(i k_i
x^i\right).
\end{equation}
Then, for two functions $f$ and $h$ the convolution theorem gives
\begin{equation}
\label{conv} \left[f~ h\right]_\mathbf{k} = \int d^3 \lambda
f_\mathbf{p} h_\mathbf{q}
\end{equation}
where $d^3
\lambda\equiv\frac{d^3pd^3q}{(2\pi)^3}\delta^3(k^i-p^i-q^i)$.

The second order Klein-Gordon equations in closed form have been
obtained in ref. \cite{Malik}. Here we use their simplified
slow-roll form from ref. \cite{Malik} (with some additional terms)
as a starting point:

\begin{eqnarray}
\delta\varphi_{I(2)\mathbf{k}}'' &+& 2\mathcal{H} \delta
\varphi_{I(2)\mathbf{k}}' +k^2\delta\varphi_{I(2)\mathbf{k}} +
\sum_K \left(a^2 U_{,\varphi_K \varphi_I}-{24 \pi G}
\varphi_{I(0)}' \varphi_{K(0)}'\right) \delta \varphi_{K(2)
\mathbf{k}} \nonumber \\
&=& - \int d^3 \lambda \left\{ a^2\sum_{K,L} \left( U_{,\varphi_I
\varphi_K \varphi_L} + \frac{8\pi G} {\mathcal{H}}\varphi_{0I}'
U_{,\varphi_K\varphi_L}\right) \delta\varphi_{K(1)\mathbf{p}}
\delta\varphi_{L(1)\mathbf{q}}\right. \nonumber \\
&+&\left(\frac{8\pi G}{\mathcal{H}}\right)^2 \sum_K\varphi_{K(0)}'
\delta\varphi_{K(1)\mathbf{p}} \left( a^2 U_{,\varphi_I}\sum_K
\varphi_{K(0)}'\delta\varphi_{K(1)\mathbf{q}} +\varphi_{I(0)}'
\sum_K a^2 U_{,\varphi_K}\delta\varphi_{K(1)\mathbf{q}}
\right)  \nonumber \\
&+&\left.\frac{16\pi G}{\mathcal{H}} a^2 \sum_K\varphi_{K(0)}'
\delta \varphi_{K(1)\mathbf{p}}\sum_K U_{,\varphi_K\varphi_I}
\delta \varphi_{K(1)\mathbf{q}} \right\} \nonumber \\
&-& \frac{8\pi G}{\mathcal{H}} \int d^3 \lambda  \left\{
2\frac{p_l q^l}{q^2}\delta\varphi_{I(1)\mathbf{p}}' \sum_K
\varphi_{K(0)}'\delta\varphi_{K(1)\mathbf{q}}' +2p^2\delta
\varphi_{I(1)\mathbf{p}} \sum_K \varphi_{K(0)}' \delta
\varphi_{K(1)\mathbf{q}} \right. \nonumber \\
\label{KGN2}&+& \left. \varphi_{I(0)}' \sum_K \left(
\left(\frac{p_lq^l +p^2}{k^2}q^2 -\frac{p_lq^l}{2}\right)
\delta\varphi_{K(1)\mathbf{p}}\delta\varphi_{K(1)\mathbf{q}}
+\left(\frac{1}{2}-\frac{q^2+p_lq^l}{k^2}\right)
\delta\varphi_{K(1)\mathbf{p}}'\delta\varphi_{K(1)\mathbf{q}}'\right)
\right\}.
\end{eqnarray}

These  equations can be simplified at small $ k $ ($ k \ll
\mathcal {H} $), insofar as in this case it is possible to
consider only the long-wavelength field perturbations at right
hand side. Indeed, the classical field perturbations was generated
from the quantum field fluctuations at the horizon crossing
\cite{LL930319}, \cite{LLbook}, \cite{Riotto}. Hence, at $ p \gg
\mathcal {H} $, $ q \gg \mathcal{H} $, there are negligibly small
quantum fluctuations instead of the classical perturbations. As a
result, (\ref{KGN2}) is reduced to
\begin{eqnarray}
\delta\varphi_{I(2)\mathbf{k}}'' &+&
2\mathcal{H}\delta\varphi_{I(2)\mathbf{k}}' +\sum_K\left(a^2
U_{,\varphi_K \varphi_I}-{24 \pi
G}\varphi_{I(0)}'\varphi_{K(0)}'\right) \delta
\varphi_{K(2)\mathbf{k}} \nonumber \\
&=& - \int d^3 \lambda \left\{ a^2\sum_{K,L} \left( U_{,\varphi_I
\varphi_K \varphi_L} + \frac{8\pi G} {\mathcal{H}}\varphi_{0I}'
U_{,\varphi_K\varphi_L}\right) \delta\varphi_{K(1)\mathbf{p}}
\delta\varphi_{L(1)\mathbf{q}}\right. \nonumber \\
&+&\left(\frac{8\pi G}{\mathcal{H}}\right)^2 \sum_K\varphi_{K(0)}'
\delta\varphi_{K(1)\mathbf{p}} \left( a^2 U_{,\varphi_I}\sum_K
\varphi_{K(0)}'\delta\varphi_{K(1)\mathbf{q}} +\varphi_{I(0)}'
\sum_K a^2 U_{,\varphi_K}\delta\varphi_{K(1)\mathbf{q}} \right)
\nonumber \\
&+&\left.\frac{16\pi G}{\mathcal{H}} a^2 \sum_K\varphi_{K(0)}'
\delta \varphi_{K(1)\mathbf{p}}\sum_K U_{,\varphi_K\varphi_I}
\delta \varphi_{K(1)\mathbf{q}} \right\}
\end{eqnarray}

One can now go back to the coordinate space. Using equation
(\ref{slow2}) and the background equations (\ref{slowbasic1}),
(\ref{slowbasic2}) for two canonical scalar fields $\varphi$,
$\chi$ on large scales we get
\begin{eqnarray}
\label{KG1phi}\delta\varphi_{(2)}' +
\frac{a^2}{3\mathcal{H}}\left( U_{,\varphi \varphi} -
\frac{U_{,\varphi} ^2 }{U_{(0)}} \right) \delta
\varphi_{(2)}+\frac{a^2}{3\mathcal{H}}\left( U_{,\chi \varphi} -
\frac{U_{,\varphi} U_{,\chi} }{U_{(0)}}\right) \delta \chi_{(2)}
&=& f_\varphi, \\
\label{KG1chi}\delta\chi_{(2)}' + \frac{a^2}{3\mathcal{H}}\left(
U_{,\chi\chi} - \frac{U_{,\chi} ^2 }{U_{(0)}} \right) \delta
\chi_{(2)}+\frac{a^2}{3\mathcal{H}}\left( U_{,\chi \varphi} -
\frac{U_{,\varphi} U_{,\chi} }{U_{(0)}}\right) \delta
\varphi_{(2)} &=& f_\chi,
\end{eqnarray}
where $ f_1 $ and $ f_2 $ are
\begin{eqnarray}
f_\varphi &=& -   \frac{a^2}{3\mathcal{H}}\left\{ \left(
U_{,\varphi \varphi \varphi} - 3\frac{U_{,\varphi}}{U_{(0)}}
U_{,\varphi \varphi}+ 2\frac{U_{,\varphi} ^3}{U_{(0)}^2}\right)
\delta\varphi_{(1)}^2 \right. \nonumber \\
&+& \left( 2U_{,\varphi \varphi \chi} - 4\frac{U_{,\varphi}
}{U_{(0)}} U_{,\varphi \chi}+ 4\frac{U_{,\varphi}^2 }{U_{(0)}
^2}U_{,\chi}-2 U_{,\varphi\varphi} \frac{U_{,\chi} }{U_{(0)}}
\right) \delta\varphi_{(1)}\delta \chi_{(1)}  \nonumber \\
&+& \left.\left( U_{,\varphi \chi \chi} - \frac{U_{,\varphi}}{
U_{(0)}} U_{,\chi \chi}- 2  U_{,\chi\varphi} \frac{U_{,\chi}}{
U_{(0)}}+ 2U_{,\varphi} \frac{U_{,\chi}^2}{U_{(0)}^2}\right)
\delta\chi_{(1)} ^2\right\},
\end{eqnarray}
\begin{eqnarray}
f_\chi &=& -   \frac{a^2}{3\mathcal{H}}\left\{ \left( U_{,\chi
\chi\chi} - 3\frac{U_{,\chi}}{U_{(0)}} U_{,\chi\chi}+ 2\frac{U_{,
\chi} ^3}{U_{(0)}^2}\right) \delta\chi_{(1)}^2
\right. \nonumber \\
&+& \left( 2U_{,\chi\chi\varphi} - 4\frac{U_{,\chi}}{U_{(0)}}
U_{,\varphi \chi}+ 4\frac{U_{,\chi}^2}{U_{(0)}^2}U_{,\varphi}-2
U_{,\chi\chi}\frac{U_{,\varphi}}{U_{(0)}} \right) \delta\chi_{(1)}
\delta \varphi_{(1)}  \nonumber \\
&+& \left. \left( U_{,\chi\varphi\varphi} - \frac{U_{,\chi}}{
U_{(0)}} U_{,\varphi\varphi}- 2  U_{,\chi\varphi} \frac{U_{,
\varphi }}{U_{(0)}}+ 2U_{,\chi} \frac{U_{,\varphi}^2}{U_{(0)}^2}
\right) \delta\varphi_{(1)} ^2\right\}.
\end{eqnarray}

\subsection{Slow-roll perturbed Einstein equations.}

Under the slow-roll condition, the equation (\ref{first0i}) can be
rewritten as
\begin{equation}
\label{first0islow}\phi_{(1)} = -\frac{1}{2}
\frac{U_{,\varphi}}{U_{(0)}} \delta \varphi_{(1)} -\frac{1}{2}
\frac{U_{,\chi}}{U_{(0)}} \delta \chi_{(1)},
\end{equation}
and the equation (\ref{firstijtr}) takes the form
\begin{eqnarray}
\phi_{(1)}' &=& \frac{4\pi G}{\mathcal{H}}
\varphi_{(0)}'\delta\varphi_{(1)}' + \frac{4\pi
G}{\mathcal{H}}\left(-a^2
\frac{U_{,\varphi\varphi}\varphi_{(0)}'+U_{,\varphi\chi}
\chi_{(0)}'}{ 3\mathcal{H}} +3 \mathcal{H} \varphi_{(0)}' \frac{
{\varphi_{(0)}'}^2+{\chi_{(0)}'}^2}{a^2U_{(0)}} \right)
\delta \varphi_{(1)} \nonumber \\
&+& \frac{4\pi G}{\mathcal{H}} \chi_{(0)}'
\delta\chi_{(1)}'+\frac{4\pi G}{\mathcal{H}}\left(-a^2 \frac{U
_{,\chi\chi} \chi_{(0)}' + U_{,\varphi\chi} \varphi_{(0)}'}{
3\mathcal{H}} +3 \mathcal{H} \chi_{(0)}' \frac{ {\varphi_{(0)}'}^2
+{\chi_{(0)}'}^2}{a^2U_{(0)}} \right) \delta\chi _{(1)}.
\end{eqnarray}

It is convenient to write the second-order perturbed  equations in
the Fourier space. At slow-roll leading order, the equation
(\ref{first00}) gives
\begin{equation}
k^2B_{(1)\mathbf{k}} =  \frac{ U_{,\varphi }}{6\mathcal{H}U_{(0)}}
\left(a^2\frac{ U_{,\varphi }^2 +U_{,\chi }^2}{ 2U_{(0)}}\delta
\varphi_{(1)\mathbf{k}} - 3\mathcal{H}\delta \varphi_{(1)
\mathbf{k}}' \right) +\frac{U_{,\chi}}{6\mathcal{H}U_{(0)}} \left(
a^2\frac{ U_{,\varphi }^2 +U_{,\chi }^2}{ 2U_{(0)}}\delta\chi_{(1)
\mathbf{k}} - 3\mathcal{H}\delta\chi_{(1) \mathbf{k}}'\right)  .
\end{equation}

The equation (\ref{second0imod}) takes the form
\begin{eqnarray}
\phi_{(2)\mathbf{k}}- \int d^3 \lambda \left\{\mathbf{p}
\mathbf{q} B_{(1) \mathbf{p}} B_{(1) \mathbf{q}} \right\} = -
\frac{1}{2} \frac{U_{,\varphi}}{U_{(0)}} \delta \varphi_{(2)
\mathbf{k}} - \frac{1}{2} \frac{U_{,\chi}}{U_{(0)}}\delta
\chi_{(2) \mathbf{k}} + \int d^3 \lambda \Bigg\{2\phi_{(1)
\mathbf{p}}\phi_{(1)\mathbf{q}}\nonumber \\
+ \frac{(\mathbf{p}\mathbf{q})^2 - p^2q^2 }{k^2q^2} \frac{
\phi_{(1) \mathbf{p}} }{6\mathcal{H}^2} \left[\frac{ U_{,\varphi
}}{U_{(0)}} \left( a^2\frac{ U_{,\varphi }^2 +U_{,\chi }^2}{
2U_{(0)}} \delta \varphi_{(1)\mathbf{q}} - 3\mathcal{H}\delta
\varphi_{(1) \mathbf{q}}'  \right) +\frac{U_{,\chi}}{U_{(0)}}
\left(a^2\frac{ U_{, \varphi }^2 +U_{,\chi }^2}{ 2U_{(0)}}\delta
\chi_{(1)\mathbf{q}} -
3\mathcal{H} \delta\chi_{(1)\mathbf{q}}' \right) \right] \nonumber \\
\label{second0islow}+3\mathcal{H} \frac{1}{a^2U_{(0)}} \left(
\frac{q^2+ \mathbf{p} \mathbf{q}
}{k^2}\delta\varphi_{(1)\mathbf{p}}' \delta\varphi_{(1)\mathbf{q}}
+ \frac{q^2+\mathbf{p} \mathbf{q}}{k^2} \delta
\chi_{(1)\mathbf{p}}' \delta \chi_{(1)\mathbf{q}} \right)
\Bigg\}.
\end{eqnarray}

\subsection{Solution of the slow-roll Klein-Gordon equations
on large scales.}

The system of the slow-roll Klein-Gordon equations can be solved
using the approach of \cite{MukhSte}. We turn now to the new
variables
\begin{equation}
x=\frac{U_{(0)}}{U_{,\varphi}}\delta\varphi_{(2)},~~~~~~
y=\frac{U_{(0)}}{U_{,\chi}}\delta\chi_{(2)}.
\end{equation}
The equations (\ref{KG1phi}), (\ref{KG1chi}) can be written as
\begin{eqnarray}
\label{KG1phixy}x' - \left( \ln\left( \frac{U_{,\varphi}}{U_{(0)}}
\right) \right) _{,\chi}\chi' \left( y - x\right) &=&
\frac{U_{(0)}}{ U_{, \varphi}}f_\varphi, \\
\label{KG1chixy}y' +\left( \ln\left(
\frac{U_{,\chi}}{U_{(0)}}\right) \right) _{, \varphi} \varphi'
\left( y - x\right) &=& \frac{U_{(0)}}{U_{,\chi}} f_\chi.
\end{eqnarray}

Their consequence
\begin{equation}
\left(x - y\right)' +  \left[ \left( \ln\left( \frac{U_{,
\varphi}}{U_{(0)}} \right) \right) _{,\chi}\chi'+\left( \ln\left(
\frac{U_{,\chi}}{U_{(0)}}\right) \right) _{, \varphi} \varphi'
\right] \left(x - y\right) =\frac{U_{(0)}}{ U_{,\varphi}}f_\varphi
-\frac{U_{(0)}}{U_{,\chi}} f_\chi
\end{equation}
has a formal solution
\begin{equation}
\label{solxy2}x - y= e^{F(\tau)}\int_{\tau_0}^{\tau}\left(
\frac{U_{(0)}}{ U_{,\varphi}}f_\varphi -\frac{U_{(0)}}{U_{,\chi}}
f_\chi\right) e^{-F(\tau')}d \tilde{\tau} + \gamma_{(2)}
e^{F(\tau)},
\end{equation}
where $\gamma_{(2)}$ is a constant and
\begin{equation}
e^{F} \equiv \exp \left\{-\int_{\tau_0}^{\tau}\left[\left(
\ln\left( \frac{U_{,\varphi}}{U_{(0)}} \right) \right)
_{,\chi}\chi' + \left( \ln\left( \frac{U_{, \chi}}{U_{(0)}}
\right) \right) _{,\varphi}\varphi' \right]d\tau \right\}.
\end{equation}

Here $\tau_0$ is an arbitrary moment of conformal time. We take
$\tau_0$ to be the moment of horizon crossing $\tau_*$ when $k =
\mathcal{H}$ for the given mode $\mathbf{k}$.

The equations (\ref{KG1phixy}), (\ref{KG1chixy}) can be integrated
now to give
\begin{eqnarray}
\label{solx2}x &=&
\int_{\tau_*}^{\tau}\left[\frac{U_{(0)}}{U_{,\varphi}}f_\varphi
-\left( \ln\left( \frac{U_{,\varphi}}{U_{(0)}}\right) \right)
_{,\chi}\chi' e^{F}\left(J+ \gamma_{(2)}\right)\right]d
\tilde{\tau}  + \alpha_{(2)}, \\
\label{soly2}y &=&
\int_{\tau_*}^{\tau}\left[\frac{U_{(0)}}{U_{,\chi}}f_\chi +\left(
\ln\left( \frac{U_{,\chi}}{U_{(0)}}\right) \right)
_{,\varphi}\varphi' e^{F}\left(J+ \gamma_{(2)}\right)\right]d \tau
' + \beta_{(2)},
\end{eqnarray}
where $\alpha_{(2)}$, $\beta_{(2)}$ are constants and
\begin{equation}
J(\tau)=\int_{\tau_*}^{\tau}\left( \frac{U_{(0)}}{
U_{,\varphi}}f_\varphi -\frac{U_{(0)}}{U_{,\chi}}
f_\chi\right)e^{-F(\tilde{\tau})}d \tilde{\tau}.
\end{equation}
Hence, the second order longwavelength fields perturbations are
\begin{eqnarray}
\label{solphi2}\delta\varphi_{(2)} &=& \frac{U_{,\varphi}
}{U_{(0)}} \int_{\tau_*} ^{\tau}\left[ \frac{U_{(0)} }{U_{,
\varphi}}f_\varphi -\left( \ln\left( \frac{U_{,\varphi}}{U_{(0)}}
\right) \right) _{,\chi}\chi' e^{F(\tau)} \left(J+ \gamma_{(2)}
\right)\right]d\tau '' + \frac{U_{,\varphi}}{U_{(0)}} \alpha_{(2)}, \\
\label{solchi2}\delta\chi_{(2)} &=&\frac{U_{,\chi}}{U_{(0)}}
\int_{\tau_*}^{\tau} \left[\frac{U_{(0)}}{U_{,\chi}}f_\chi +\left(
\ln\left( \frac{ U_{,\chi} }{U_{(0)}} \right) \right)
_{,\varphi}\varphi' e^{F(\tau)}\left( J+ \gamma_{(2)}
\right)\right]d \tau ' + \frac{U_{,\chi}}{U_{(0)}}\beta_{(2)}.
\end{eqnarray}

The consideration of equations (\ref{solx2}),(\ref{soly2}) and
(\ref{solxy2}) at $ \tau = \tau_* $ gives
\begin{equation}
x_* - y_*= \gamma_{(2)} e^{F(\tau_*)}= \gamma_{(2)}=\alpha_{(2)} -
\beta_{(2)}.
\end{equation}

Similarly, we get the first order longwavelength slow-roll
solution
\begin{eqnarray}
\label{solphi1}\delta\varphi_{(1)}&=& -\gamma_{(1)} \frac{U_{,
\varphi}}{U_{(0)}} \int_{\tau_*} ^{\tau} \left( \ln\left(
\frac{U_{, \varphi}}{U_{(0)}} \right) \right) _{,\chi}\chi'
e^{F(\tau)}d \tau '' +\frac{ U_{,\varphi}}{U_{(0)}}
\alpha_{(1)},\\
\label{solchi1}\delta \chi_{(1)} &=& \gamma_{(1)} \frac{U_{,
\chi}}{U_{(0)}} \int_{\tau_*} ^{\tau}\left( \ln\left( \frac{U_{,
\chi}}{U_{(0)}} \right) \right) _{,\varphi}\varphi' e^{F(\tau)}d
\tau ' + \frac{U_{,\chi}}{U_{(0)}} \beta_{(1)},
\end{eqnarray}
where $\alpha_{(1)}$, $\beta_{(1)}$, $\gamma_{(1)}$ are constants
constrained  by
\begin{equation}
\gamma_{(1)}=\alpha_{(1)} - \beta_{(1)}.
\end{equation}

\subsection{Curvature perturbation in slow-roll.}

At leading order in slow-roll parameters, we have
\begin{equation}
\label{rho1}\delta\rho_{(1)} =  U_{,\varphi} \delta\varphi_{(1)}+
U_{,\chi} \delta\chi_{(1)},
\end{equation}
\begin{equation}
\label{rho1prime}\delta\rho_{(1)}' = \left(U_{,\varphi\varphi}
\varphi_{(0)}' +U_{,\varphi\chi} \chi_{(0)}'
\right)\delta\varphi_{(1)} + U_{,\varphi} \delta\varphi_{(1)}'+
\left(U_{,\chi\chi}\chi' +U_{,\chi\varphi}\varphi'
\right)\delta\chi_{(1)}+ U_{,\chi} \delta\chi_{(1)}'.
\end{equation}

Using the background slow-roll equations, the first order
equations (\ref{phi1slow}), (\ref{chi1slow}) and the Einstein
equations (\ref{first0islow}) - (\ref{second0islow}), we obtain a
simple expression
\begin{equation}
\delta\rho_{(2)\mathbf{k}} =  U_{ ,\varphi} \delta\varphi_{(2)
\mathbf{k}}  + U_{,\chi} \delta \chi_{(2)\mathbf{k}}  + \int d^3
\lambda \left\{U_{, \varphi \varphi} \delta \varphi_{(1)
\mathbf{p}} \delta \varphi _{(1) \mathbf{q}} + U_{,\chi\chi}
\delta \chi_{(1)\mathbf{p}} \delta \chi_{(1)\mathbf{q}} +
2U_{,\varphi\chi} \delta \varphi_{(1)\mathbf{p}} \delta \chi_{(1)
\mathbf{q}} \right\}
\end{equation}
or
\begin{equation}
\label{rho2}\delta\rho_{(2)} =  U_{ ,\varphi} \delta\varphi_{(2)}
+ U_{,\chi} \delta \chi_{(2)}  + U_{,\varphi\varphi} \delta
\varphi_{(1)} ^2 + U_{,\chi\chi} \delta \chi_{(1)} ^2 +
2U_{,\varphi\chi} \delta \varphi_{(1)} \delta\chi_{(1)} .
\end{equation}

The substitution of equations (\ref{rho1}), (\ref{rho1prime}),
(\ref{rho2}) into (\ref{zeta2UC}) gives
\begin{eqnarray}
\zeta_{(2)} &=&\frac{8\pi G U_{(0)}}{U_{,\varphi}^2+U_{,\chi}^2}
\bigg\{
U_{,\varphi}\delta \varphi_{(2)} +U_{,\chi} \delta \chi_{(2)} \nonumber \\
&-& \left(U_{,\varphi\varphi}-\frac{U_{,\varphi}^2}{U_{(0)}} -4
\frac{ U_{,\varphi\varphi} U_{,\chi} -U_{,\varphi\chi} U_{,
\varphi} }{ U_{,\varphi}^2 +U_{,\chi}^2 }U_{,\chi} -2 \frac{
U_{,\varphi \varphi} U_{,\varphi}^2+2 U_{,\varphi\chi} U_{,
\varphi} U_{,\chi} +U_{,\chi\chi}U_{,\chi}^2}{ ( U_{,\varphi}^2
+U_{,\chi}^2)^2}U_{,\varphi}^2\right) \delta \varphi_{(1)}^2 \nonumber \\
&-& \left(U_{,\chi\chi}-\frac{U_{,\chi}^2}{U_{(0)}} -4 \frac{
U_{,\chi\chi} U_{,\chi} -U_{,\varphi\chi}U_{,\chi}}{
U_{,\varphi}^2 +U_{,\chi}^2}U_{,\varphi} -2 \frac{ U_{,\chi\chi}
U_{,\chi}^2+2 U_{,\varphi\chi} U_{,\varphi}U_{,\chi}
+U_{,\varphi\varphi} U_{,\varphi}^2}{ ( U_{,\varphi}^2
+U_{,\chi}^2) ^2}U_{,\chi}^2\right)\delta \chi_{(1)}^2 \nonumber \\
\label{zeta2slow} &-& 2\left(U_{,\varphi\chi} -\frac{U_{,\varphi}
U_{, \chi}}{U_{(0)}} -2 \frac{ 2U_{,\varphi\chi}U_{,\varphi}
U_{,\chi}-U_{,\varphi\varphi} U_{,\chi}^2 -U_{,\chi\chi} U_{,
\varphi} ^2}{ ( U_{,\varphi}^2 +U_{,\chi}^2) ^2}U_{,\varphi}
U_{,\chi}\right) \delta \varphi_{(1)} \delta \chi_{(1)}\bigg\}.
\end{eqnarray}
where formal expression for $ \delta \varphi_{(2)} $ and $ \delta
\chi_ {(2)} $ are given by the equations (\ref{solphi2}) and
(\ref{solchi2}).

This equation is the main result of this work. It is follows from
(\ref{zeta2slow}), that the first order fields perturbations give
a local contribution to $ \zeta_{(2)} $, both directly and
implicitly through the quantities $ \delta \varphi_{(2)} $, $
\delta \chi_{(2)} $.

\subsubsection{Product potential, $U(\varphi,\chi) =
V(\varphi)W(\chi)$.}

The formal expressions (\ref{solphi1}),(\ref{solchi1}) and
(\ref{solphi2}), (\ref{solchi2}) for product potential can be
integrated to  give correspondingly
\begin{equation}
\label{prod1}\delta \varphi_{(1)}  = \frac {V _ {,\varphi}}
{V_{(0)}} \frac{V_{(0)*}} {V _ {,\varphi *}}\delta \varphi_{(1)*}
, ~~~~\delta \chi_{(1)} = \frac{W _ {,\chi}} {W_{(0)}}
\frac{W_{(0)*}}{W _ {,\chi *}} \delta \chi_{(1)*} ,
\end{equation}
and
\begin{eqnarray}
\label{prodphi2}\delta \varphi_{(2)}(\tau)  &=& \left( \left.\frac
{V_{,\varphi \varphi }} {V}\right|^\tau _{\tau_*} - \left.
\frac{V_{,\varphi }^2}{V^2}\right|^\tau _{\tau_*} \right)
\frac{V_{,\varphi}(\tau) }{V_{(0)}(\tau)} \frac{V_{(0)*}^2}{V _
{,\varphi *}^2} \delta \varphi _{(1)*}^2
+\frac{V_{,\varphi}(\tau)}{V_{(0)}(\tau)} \frac{V_{(0)
*}}{V_{,\varphi *}} \delta \varphi_{(2)*}   , \\
\label{prodchi2}\delta \chi_{(2)}(\tau) &=& \left( \left.\frac
{W_{,\chi\chi}} {W}\right|^\tau _{\tau_*} - \left. \frac{W_{,\chi
}^2}{W^2} \right|^\tau _{\tau_*} \right) \frac{W_{,\chi}(\tau)}{
W_{(0) }(\tau)}\frac{W_{(0)*}^2}{W _ {,\chi *}^2} \delta
\chi_{(1)*}^2 + \frac{W_{,\chi}(\tau) }{W_{(0)}(\tau)}
\frac{W_{(0)*}}{W _ {,\varphi *}}\delta \chi_{(2)*}
\end{eqnarray}

Using the background slow-roll equations and equations
(\ref{prod1}), (\ref{prodphi2}), (\ref{prodphi2}), the equation
(\ref{zeta2slow}) can be simplified to give
\begin{eqnarray}
\zeta_{(2)} &=&  \frac{8\pi G V_{,\varphi}^2W_{(0)}^2}{
V_{,\varphi}^2W_{(0)}^2 + V_{(0)}^2W_{,\chi}^2} \frac{V_{(0)*}}{V
_ {,\varphi *}}\delta \varphi _{(2)*}  + \frac{8\pi G V_{(0)}^2
W_{,\chi}^2}{ V_{,\varphi}^2W_{(0)}^2+V_{(0)}^2 W_{,\chi}^2}
\frac{W_{(0)*}}{W _ {,\varphi *}}\delta \chi_{(2)*} \nonumber \\
&-& \frac{8\pi G V_{,\varphi}^2W_{(0)}^2}{V_{,\varphi}^2W_{(0)}^2
+V_{(0)}^2W_{,\chi}^2}\left(  \frac {V_{,\varphi \varphi *}}
{V_{(0)*}} - \frac{V_{,\varphi *}^2}{V_{(0)*}^2} + 2\frac{
2V_{,\varphi}^2 W_{,\chi}^2 -W_{,\chi}^2V_{(0)} V_{,\varphi
\varphi} - V_{,\varphi}^2W_{(0)} W_{,\chi\chi} }{
(V_{,\varphi}^2W^2 +V^2W_{,\chi}^2)^2} V_{(0)}^2W_{,\chi}^2\right)
\frac{V_{(0)*}^2}{V _ {,\varphi *}^2} \delta \varphi_{(1)*}^2  \nonumber \\
&-& \frac{8\pi G V_{(0)}^2W_{,\chi}^2}{V_{,\varphi}^2W_{(0)}^2
+V_{(0)}^2W_{,\chi}^2}\left(\frac {W_{,\chi\chi*}} {W_{(0)*}} -
\frac{W_{,\chi *}^2}{W_{(0)*}^2} + 2\frac{ 2V_{,\varphi}^2
W_{,\chi}^2 -W_{,\chi}^2V_{(0)} V_{,\varphi \varphi} -
V_{,\varphi}^2W_{(0)} W_{,\chi\chi} }{ (V_{,\varphi}^2W_{(0)}^2
+V_{(0)}^2W_{,\chi}^2)^2} W_{(0)}^2V_{,\varphi}^2\right)
\frac{W_{(0)*}^2}{W _{,\chi *}^2} \delta \chi_{(1)*}^2    \nonumber  \\
\label{zeta2prod}  &+& 32\pi G \frac{  V_{,\varphi}^2W_{,\chi}^2
V_{(0)}^2W_{(0)}^2}{ V_{,\varphi}^2W_{(0)}^2
+V_{(0)}^2W_{,\chi}^2} \frac{
2V_{,\varphi}^2W_{,\chi}^2-V_{,\varphi\varphi} W_{,\chi}^2V_{(0)}
-W_{,\chi\chi} V_{,\varphi}^2W_{(0)}}{ (V_{,\varphi}^2W_{(0)}^2
+V_{(0)}^2W_{,\chi}^2) ^2} \frac{V_{(0)*}}{V _ {,\varphi
*}}\frac{W_{(0)*}}{W _ {,\chi *}} \delta \chi_{(1)*} \delta
\varphi_{(1)*}.
\end{eqnarray}

This result is in agreement with ref. \cite{0701247}, where was
considered two scalar fields with non-canonical kinetic terms.

\subsubsection{Sum potential, $U=V(\varphi) + W(\chi)$.}
At first order we obtain the known result \cite{MukhSte}
\begin{equation}
\delta\varphi_{(1)}=V_{,\varphi}\frac{C-DW_{(0)}}{U_{(0)}},
~~~~\delta\chi_{(1)}=W_{,\chi}\frac{C+DV_{(0)}}{U_{(0)}}
\end{equation}
where
\begin{equation}
C = \frac{V_{(0)*} }{V_{,\varphi *}} \delta\varphi_{(1)*}+ \frac{
W_{(0)*}}{W_{,\chi *}}\delta\chi_{(1)*} ,~~~~~~ D=\frac{\delta
\chi_{(1)*}} {W_{,\chi *}} - \frac{\delta \varphi_{(1)*}
}{V_{,\varphi *}}  .
\end{equation}
The second order equations (\ref{solphi2}), (\ref{solchi2}) give
\begin{eqnarray}
\frac{U_{(0)}}{V_{,\varphi}} \delta\varphi_{(2)} &=& C^2\left.
\left( \frac{ V_{,\varphi \varphi}}{U_{(0)}} - \frac{ V_{,\varphi
}^2}{U_{(0)}^2} - \frac{ W_{,\chi }^2}{U_{(0)}^2}\right) \right|
_{\tau_*}^\tau + D^2\left.\left(\frac{V_{,\varphi \varphi}
W_{(0)}^2}{U_{(0)}} -\frac{ V_{,\varphi} ^2W_{(0)}^2 }{U_{(0)}^2}
- \frac{W_{,\chi}^2V_{(0)}^2}{U_{(0)}^2}\right) \right|_{\tau_*}
^\tau \nonumber \\
&-& 2CD\left.\left( \frac{V_{,\varphi \varphi} W_{(0)}}{U_{(0)}} -
\frac{ V_{,\varphi}^2W_{(0)}}{U_{(0)}^2} +\frac{ W_{,\chi}^2
V_{(0)}}{U_{(0)} ^2} \right) \right|_{\tau_*}^\tau
+\tilde{\gamma}_{(2)}\frac{W_{(0)}-W_{(0)*}}{U_{(0)*}}+\alpha_{(2)}, \\
\frac{U_{(0)}}{W_{,\chi}} \delta\chi_{(2)} &=&  C^2\left.\left(
\frac{ W_{,\chi\chi}}{U_{(0)}} - \frac{ W_{,\chi }^2}{U_{(0)}^2} -
\frac{ V_{,\varphi }^2}{U_{(0)}^2}\right)\right|_{\tau_*}^\tau +
D^2\left. \left( \frac{W_{,\chi\chi}V_{(0)}^2}{U_{(0)}} -\frac{
W_{,\chi} ^2V_{(0)}^2}{U_{(0)}^2} - \frac{V_{,\varphi}^2
W_{(0)}^2}{U_{(0)}^2}\right) \right|_{\tau_*}^\tau \nonumber \\
&+& 2CD\left.\left( \frac{W_{,\chi\chi} V_{(0)}}{U_{(0)}} - \frac{
W_{,\chi}^2V_{(0)}}{U_{(0)}^2} +\frac{
V_{,\varphi}^2W_{(0)}}{U_{(0)}^2}\right) \right|_{\tau_*}^\tau
-\tilde{\gamma}_{(2)}\frac{V_{(0)}-V_{(0)*}}{U_{(0)*}}+\beta_{(2)},
\end{eqnarray}
where
\begin{equation}
\tilde{\gamma}_{(2)} =\gamma_{(2)} -C^2\left( \frac{ V_{,\varphi
\varphi *} -W_{,\chi\chi *}}{U_{(0)*}} \right)-D^2\left(\frac{
V_{,\varphi \varphi *}W_{(0)*}^2}{U_{(0)*}}- \frac{ W_{,\chi\chi
*}V_{(0)*}^2}{U_{(0)*}}\right)+ 2CD\left( \frac{V_{,\varphi
\varphi *} W_{(0)*}}{U_{(0)*}} +\frac{W_{,\chi\chi*} V_{(0)*}
}{U_{(0)*}}\right).
\end{equation}

The second order curvature perturbation $ \zeta^{(2)} $ is
\begin{eqnarray}
\zeta_{(2)} &=&  -\frac{8\pi G }{V_{,\varphi}^2+W_{,\chi}^2}
\Bigg\{ \Bigg(  V_{,\varphi \varphi *} \left( V_{,\varphi}^2
(W_{(0)}+V_{(0)*}) -W_{,\chi} ^2(V_{(0)}-V_{(0)*})\right) -
V_{,\varphi *}^2 \left(V_{,\varphi}^2 +W_{,\chi}^2 \right)
\nonumber \\
&+& V_{,\varphi}^2 W_{,\chi}^2\left( 1 -2 \frac{ V_{,\varphi
\varphi} W_{,\chi}^2 +W_{,\chi\chi} V_{,\varphi} ^2}{ (V_{,
\varphi} ^2+W_{,\chi}^2)^2} U_{(0)}\right)\Bigg) \frac{\delta
\varphi_{(1)*}^2}{V_{,\varphi *}^2} + \Bigg( V_{,\varphi}^2
W_{,\chi}^2\left( 1 -2 \frac{ V_{,\varphi \varphi} W_{,\chi}^2
+W_{,\chi\chi} V_{,\varphi} ^2}{ (V_{,\varphi}^2+W_{,\chi}^2)^2}
U_{(0)}\right)
\nonumber \\
&+& W_{,\chi\chi *} \left( W_{,\chi}^2(V_{(0)} +W_{(0)*}) -V_{,
\varphi} ^2(W_{(0)} - W_{(0)*}) \right) -  W_{,\chi *}^2 \left(
V_{,\varphi}^2 + W_{,\chi}^2\right) \Bigg)
\frac{\delta\chi_{(1)*}^2}{W_{,\chi *}^2}
\nonumber \\
&-& 2 V_{,\varphi}^2 W_{,\chi}^2\left( 1 -2 \frac{ V_{,\varphi
\varphi} W_{,\chi}^2+W_{,\chi\chi} V_{,\varphi} ^2}{
(V_{,\varphi}^2 + W_{,\chi}^2)^2} U_{(0)}\right) \frac{\delta
\varphi_{(1)*} \delta\chi_{(1)*}}{ V_{,\varphi *} W_{,\chi *}}
\Bigg\} \nonumber \\
\label{zeta2sum}&+& 8\pi G \left(V_{(0)*} -\frac{W_{,\chi}
^2V_{(0)}-V_{, \varphi}^2 W_{(0)}}{ V_{,\varphi }^2 +W_{,\chi}^2}
\right) \frac{\delta\varphi_{(2)*}}{ V_{, \varphi *}} +8\pi G
\left( W_{(0)*} - \frac{V_{,\varphi}^2W_{(0)} -W_{,\chi}^2
V_{(0)}}{V_{,\varphi}^2 +W_{,\chi}^2}  \right) \frac{ \delta
\chi_{(2)*}}{W_{,\chi *}} .
\end{eqnarray}

This result is identical to that found in ref.
\cite{Vernizzi_Wands}.

\section{Non-Gaussianity.}
\label{fnlcec}

Let us consider the spectrum $P_\zeta(\mathbf{k})$ and bispectrum
$B_\zeta(\mathbf{k}_1,\mathbf{k}_2, \mathbf{k}_3)$, defined by
\begin{equation}
\label{2p}\langle\zeta_{\mathbf{k}_1}\zeta_{\mathbf{k}_2} \rangle
=(2\pi)^3\delta^{(3)}(\mathbf{k}_1+\mathbf{k}_2
)P_\zeta(\mathbf{k}_1),
\end{equation}
\begin{equation}
\label{3p}\langle\zeta_{\mathbf{k}_1}\zeta_{\mathbf{k}_2}\zeta
_{\mathbf{k} _3} \rangle
=(2\pi)^3\delta^{(3)}(\mathbf{k}_1+\mathbf{k}_2 +\mathbf{k}_3)
B_\zeta(\mathbf{k}_1,\mathbf{k}_2, \mathbf{k}_3).
\end{equation}

The bispectrum is parametrized in terms of products of the
spectrum and the dimensionless nonlinear parameter $f_{NL}$ as
\cite{BL}
\begin{equation}
\label{bisp}B_\zeta(\mathbf{k}_1,\mathbf{k}_2, \mathbf{k}_3)
=\frac{6}{5}f_{NL} (\mathbf{k}_1, \mathbf{k}_2, \mathbf{k}_3)
\left(P_\zeta(\mathbf{k}_1)P_\zeta(\mathbf{k}_2) +
P_\zeta(\mathbf{k}_2)P_\zeta(\mathbf{k}_3)
+P_\zeta(\mathbf{k}_3)P_\zeta(\mathbf{k}_1) \right).
\end{equation}

At leading order, we have
\begin{equation}
\label{2plead}\langle\zeta_{\mathbf{k}_1}\zeta_{\mathbf{k}_2}\rangle
=\langle\zeta_{(1)\mathbf{k}_1}\zeta_{(1)\mathbf{k}_2} \rangle
=(2\pi)^3\delta^{(3)}(\mathbf{k}_1+\mathbf{k}_2
)P_{\zeta_{(1)}}(\mathbf{k}_1,\mathbf{k}_2),
\end{equation}
\begin{equation}
\label{3plead}\langle\zeta_{\mathbf{k}_1}\zeta_{\mathbf{k}_2}\zeta
_{\mathbf{k} _3} \rangle =\frac{1}{2}\langle\zeta_{(1)
\mathbf{k}_1}\zeta_{(1) \mathbf{k}_2}\zeta _{(2)\mathbf{k} _3}
\rangle + \frac{1}{2} \langle\zeta_{(1) \mathbf{k}_1}
\zeta_{(2)\mathbf{k}_2}\zeta _{(1)\mathbf{k} _3} \rangle
+\frac{1}{2} \langle\zeta_{(2)
\mathbf{k}_1}\zeta_{(1)\mathbf{k}_2}\zeta _{(1)\mathbf{k} _3}
\rangle
\end{equation}
Since the variable $\zeta_{\mathbf{k}}$ is Gaussian, the Wick's
theorem yields
\begin{equation}
\label{zeta1_4}\langle\zeta_{(1)\mathbf{k}_1}
\zeta_{(1)\mathbf{k}_2}[\zeta _{(1)}^2]_{\mathbf{k} _3} \rangle
=2(2\pi)^3\delta^{(3)} (\mathbf{k}_1 +\mathbf{k}_2 +\mathbf{k}_3)
P_{\zeta_{(1)}} (\mathbf{k}_1)P_{\zeta_{(1)}}(\mathbf{k}_2).
\end{equation}
Combining equations (\ref{3p}) - (\ref{zeta1_4}), one can write
\begin{equation}
\label{fnlst}\frac{6}{5}f_{NL} (\mathbf{k}_1, \mathbf{k}_2,
\mathbf{k}_3) =\frac{\langle\zeta_{(1) \mathbf{k}_1} \zeta_{(1)
\mathbf{k}_2}\zeta _{(2)\mathbf{k} _3} \rangle + \mathrm{c}.
~\mathrm{p}.}{\langle \zeta_{(1)\mathbf{k}_1} \zeta_{(1)
\mathbf{k}_2} \zeta _{(1)\mathbf{k} _3}^2 \rangle+ \mathrm{c}.~
\mathrm{p}. },
\end{equation}
where $\mathrm{c}.~ \mathrm{p}.$ means cyclic permutations.

In the linear perturbation theory, all perturbations are separated
on adiabatic and entropy ones. To do this, one can introduce the
variables $\delta s_{(1)} $, $\delta \sigma_{(1)} $ \cite{GWBM}
\begin{equation}
\delta s_{(1)} = \cos \Theta \delta \chi_{(1)} - \sin \Theta
\delta \varphi_{(1)} ,~~~~~ \delta \sigma_{(1)} = \cos \Theta
\delta \varphi_{(1)} + \sin \Theta \delta \chi_{(1)} ,
\end{equation}
where
$$
\cos \Theta =\frac{\varphi_{(0)}'}{\sigma'},~~~~~\sin \Theta
=\frac{\chi_{(0)}'}{\sigma'}
$$
and
$$
\sigma' =\sqrt{\varphi_{(0)}^{\prime 2} +\chi_{(0)}^{\prime 2}}.
$$
Perturbations in $\delta \sigma_{(1)} $, with $\delta s_{(1)} =0$,
describe adiabatic field perturbations \cite{GWBM}. Perturbations
with $\delta \sigma_{(1)} =0$ represent the entropy perturbations.

Going to the non-linear case, it is very convenient to use the
equality \cite{GWBM}
\begin{equation}
\zeta_{(1)}= -\mathcal{H}\frac{\delta\sigma_{(1)}}{\sigma '}
\end{equation}
and to introduce the rescaled variables
\begin{eqnarray}
\delta \tilde{s}_{(1)} &\equiv&  \mathcal{H}\frac{\delta
s_{(1)}}{ \sigma '}, \\
\delta\tilde{\sigma}_{(1)} &\equiv& \mathcal{H} \frac{\delta
\sigma_{(1)}}{\sigma '}.
\end{eqnarray}

Using these first order variables (we do not employ the nonlinear
generalization of ref. \cite{0610064}), the second order curvature
perturbation can be parameterized as
\begin{equation}
\zeta_{(2)}= \delta\tilde{\sigma}_{(2)}+
C_{\tilde{\sigma}\tilde{\sigma}}\zeta_{(1)}^2-
2C_{\tilde{s}\tilde{\sigma}}\zeta_{(1)}\delta \tilde{s}_{(1)}+
C_{\tilde{s}\tilde{s}}\delta \tilde{s}_{(1)}^2
\end{equation}
where  $C_{\tilde{\sigma}\tilde{\sigma}}$, $C_{\tilde{s} \tilde{
\sigma}}$, $C_{\tilde{s}\tilde{s}}$ are some functions of the
background fields, and the quantity $\delta\tilde{\sigma}_2$ is
zero if the initial fields perturbations are Gaussian. Neglecting
the initial non-Gaussianity of the scalar fields $\varphi$ and
$\chi$, we obtain
\begin{equation}
\label{fnl1}\frac{6}{5}f_{NL} (\mathbf{k}_1, \mathbf{k}_2,
\mathbf{k}_3) = C_{\tilde{\sigma}\tilde{\sigma}} -2C_{\tilde{s}
\tilde{ \sigma}} \frac{\langle \zeta_{(1) \mathbf{k} _1}
\zeta_{(1)\mathbf{k}_2} [\zeta_{(1)}\delta \tilde{s}_{(1)}]
_{\mathbf{k} _3} \rangle +\mathrm{c}.~ \mathrm{p}.}{ \langle \zeta
_{(1) \mathbf{k}_1} \zeta_{(1)\mathbf{k}_2} \zeta _{(1)\mathbf{k}
_3}^2 \rangle + \mathrm{c}.~ \mathrm{p}.}
+C_{\tilde{s}\tilde{s}}\frac{\langle \zeta_{(1)
\mathbf{k}_1}\zeta_{(1)\mathbf{k}_2}[\delta \tilde{s} _{(1) }^2]
_{\mathbf{k} _3} \rangle +\mathrm{c}.~ \mathrm{p}.}{ \langle
\zeta_{(1) \mathbf{k}_1} \zeta_{(1)\mathbf{k}_2} \zeta
_{(1)\mathbf{k} _3}^2 \rangle + \mathrm{c}.~ \mathrm{p}.} .
\end{equation}

The calculation of the correlation functions yields
\begin{equation}
\frac{6}{5}f_{NL} (\mathbf{k}_1, \mathbf{k}_2, \mathbf{k}_3)
=C_{\tilde{\sigma}\tilde{\sigma}} +C_{\tilde{s}\tilde{s}}
\frac{K(\mathbf{k}_1)K(\mathbf{k}_2) + K(\mathbf{k}_2)
K(\mathbf{k}_3)+ K(\mathbf{k}_1)K(\mathbf{k}_3)}{ P_\zeta
(\mathbf{k}_1) P_\zeta(\mathbf{k}_2) + P_\zeta( \mathbf{k}_2)
P_\zeta(\mathbf{k}_3) +P_\zeta(\mathbf{k}_1)P_\zeta(\mathbf{k}_3)}
\nonumber
\end{equation}
\begin{equation}
\label{fnl}- C_{\tilde{s} \tilde{ \sigma}}
\frac{P_\zeta(\mathbf{k}_1) K(\mathbf{k}_2)
+P_\zeta(\mathbf{k}_2)K(\mathbf{k}_1) +P_\zeta(\mathbf{k}_2)
K(\mathbf{k}_3) +P_\zeta(\mathbf{k}_3)K(\mathbf{k}_2) +
P_\zeta(\mathbf{k}_1) K(\mathbf{k}_3)
+P_\zeta(\mathbf{k}_3)K(\mathbf{k}_1)
}{P_\zeta(\mathbf{k}_1)P_\zeta(\mathbf{k}_2) +
P_\zeta(\mathbf{k}_2)P_\zeta(\mathbf{k}_3)
+P_\zeta(\mathbf{k}_3)P_\zeta(\mathbf{k}_1)},
\end{equation}
where we introduce the notation
\begin{equation}
K(\mathbf{k})=\langle \zeta_{(1)\mathbf{k}} \tilde{s}_
{-\mathbf{k}} \rangle .
\end{equation}

Spectra and correlation functions are easy to calculate, using the
well-known result \cite{GWBM}
\begin{equation}
\label{initial} \delta \varphi _{(1)\mathbf{k}*} = \frac{H_*}{
\sqrt{2k^3} }e_{\varphi\mathbf{k}}, ~~~~~\delta \chi
_{(1)\mathbf{k}*}  = \frac{H_*}{\sqrt{2k^3} }e_{\chi\mathbf{k}},
\end{equation}
where $e_{\varphi\mathbf{k}}$ and $e_{\chi\mathbf{k}}$ are
independent Gaussian random variables with zero mean and unit
variance.

The analysis of the expression (\ref{fnl}) is too complicated for
general form of potential $U(\varphi,\chi)$, so let us consider
only some particular cases.

\subsection{Product potential, $U(\varphi,\chi) =
V(\varphi)W(\chi)$.}

In this case we have
\begin{eqnarray}
\label{s1prod} \delta \tilde{s}_{(1)} &=& \frac{8 \pi G
V_{(0)}W_{(0)}V_{,\varphi}W_{,\chi}}{V_{, \varphi}^2W_{(0)}^2
+V_{(0)}^2W_{,\chi}^2}\left( \frac{V_{(0)*}}{V _ {,\varphi
*}}\delta \varphi_{(1)*} - \frac{W_{(0)*}}{W _ {,\chi *}} \delta
\chi_{(1)*}\right), \\
\label{sigma1prod} \delta \tilde{\sigma} _{(1)} &=& -\frac{8 \pi G
}{V_{, \varphi}^2W_{(0)}^2 +V_{(0)}^2W_{,\chi}^2}\left(
V_{(0)}^2W_{,\chi}^2\frac{W_{(0)*}}{W _ {,\chi *}} \delta
\chi_{(1)*} + W_{(0)}^2V_{,\varphi}^2\frac{V_{(0)*}}{V _ {,\varphi
*}}\delta \varphi_{(1)*}\right) .
\end{eqnarray}
The equation (\ref{zeta2prod}) can be rewritten as
\begin{eqnarray}
\zeta_{(2)}  &=& \frac{1}{8\pi G }\left[\left( \left( \frac{
V_{,\varphi *}^2}{V_{(0)*}^2}-\frac {V_{,\varphi \varphi *}}
{V_{(0)*}} \right) \cos^2 \Theta+ \left( \frac{W_{,\chi * }^2}{
W_{(0)*}^2}-\frac {W_{,\chi \chi}*} {W_{(0)*}} \right) \sin^2
\Theta\right) \zeta_{(1)}^2 \right.\nonumber \\
&+& \left( \left( \frac{V_{,\varphi *}^2}{V_{(0)*}^2}- \frac
{V_{,\varphi \varphi *}} {V_{(0)*}} \right) \sin^2 \Theta + \left(
\frac{W_{,\chi *}^2}{W_{(0)*}^2} -\frac {W_{,\chi\chi}*}
{W_{(0)*}} \right) \cos^2 \Theta \right.\nonumber \\
&+&\left. 2\sin \Theta \cos \Theta \left(\frac{V_{,\varphi
\varphi} }{V_{,\varphi}} \frac{W_{, \chi}}{W_{(0)}}+\frac{W_{,\chi
\chi}}{W_{,\chi}}\frac{V_{,\varphi}}{V_{(0)}} - 2 \frac{V_{,
\varphi}W_{,\chi}}{V _{(0)}W_{(0)}} \right)\right)
\delta\tilde{s}_{(1)}^2 \nonumber \\
\label{zeta2prodmod}&+& \left.2\sin \Theta \cos \Theta\left( \frac
{W_{,\chi \chi}*} {W_{(0)*}} - \frac{W_{,\chi * }^2}{W_{(0)*}^2}
-\frac {V_{,\varphi \varphi *}} {V_{(0)*}} + \frac{V_{,\varphi
*}^2}{V_{(0)*}^2} \right)\delta \tilde{s}_{(1)}\zeta_{(1)}\right]
-\delta\tilde{\sigma}_{(2)},
\end{eqnarray}

where $\delta\tilde{\sigma}_{(2)}$ is obtained from
$\delta\tilde{\sigma}_{(1)}$ by replacing  $\delta
\varphi_{(1)*}$, $\delta \chi_{(1)*}$ by $\delta \varphi_{(2)*}$,
$\delta \chi_{(2)*}$.

The equation above shows that the coefficient
$C_{\tilde{\sigma}\tilde{\sigma}}$ is of the order of slow-roll
parameters, i.e.
\begin{equation}
\label{coeff} |C_{\tilde{\sigma}\tilde{\sigma}} | \ll 1.
\end{equation}
This inequality and equation (\ref{fnl1}) indicate that in models
with large $f_{NL}$ the entropy perturbation $\delta
\tilde{s}_{(1)} $ can not be neglected. In addition, it follows
now that a significant level of non-Gaussianity $ |f_ {NL}|
\gtrsim 1 $  is possible if and only if
\begin{equation}
\Bigg| C_{\tilde{s} \tilde{ \sigma}} \frac{P_\zeta(\mathbf{k}_1)
K(\mathbf{k}_2) +P_\zeta(\mathbf{k}_2)K(\mathbf{k}_1)
+P_\zeta(\mathbf{k}_2) K(\mathbf{k}_3)
+P_\zeta(\mathbf{k}_3)K(\mathbf{k}_2) + P_\zeta(\mathbf{k}_1)
K(\mathbf{k}_3) +P_\zeta(\mathbf{k}_3)K(\mathbf{k}_1)
}{P_\zeta(\mathbf{k}_1)P_\zeta(\mathbf{k}_2) +
P_\zeta(\mathbf{k}_2)P_\zeta(\mathbf{k}_3)
+P_\zeta(\mathbf{k}_3)P_\zeta(\mathbf{k}_1)} \nonumber
\end{equation}
\begin{equation}
\label{largefnl}- C_{\tilde{s}\tilde{s}}
\frac{K(\mathbf{k}_1)K(\mathbf{k}_2) + K(\mathbf{k}_2)
K(\mathbf{k}_3)+ K(\mathbf{k}_1)K(\mathbf{k}_3)}{ P_\zeta
(\mathbf{k}_1) P_\zeta(\mathbf{k}_2) + P_\zeta( \mathbf{k}_2)
P_\zeta(\mathbf{k}_3) +P_\zeta(\mathbf{k}_1)P_\zeta(\mathbf{k}_3)}
\Bigg| \gtrsim 1 .
\end{equation}

Consider the case that can be easily implemented
\begin{equation}
\label{coeff} \max \{|C_{\tilde{s} \tilde{ \sigma}}  |,
|C_{\tilde{s}\tilde{s}} |\} \ll 1.
\end{equation}
Then the mandatory condition of the presence of large
non-Gaussianity is
\begin{equation}
\label{largefnls}|K(\mathbf{k})|\gg P_\zeta(\mathbf{k}).
\end{equation}

For this class of models, large $ f_ {NL} $ is achieved when two
conditions are met: 1) the variables $\delta\tilde{\sigma}_{(1)}$
and $\delta \tilde{s} _{(1)}$ are highly correlated; 2) entropy
perturbations dominate over adiabatic ones, $ P_{\tilde{s}}
(\mathbf{k})\gg P_{\delta\tilde{\sigma}}( \mathbf{k})$.

Quantitatively, the equations (\ref{initial}), (\ref{s1prod}) and
(\ref{sigma1prod}) lead to the expressions
\begin{eqnarray}
K(\mathbf{k}) &=& (8 \pi G )^3 \sin \Theta \cos \Theta
\left(\left( \frac{W_{,\chi}}{W_{(0)}}\frac{W_{(0)*}}{W _ {,\chi
*}} \right)^2 -\left( \frac{V_{,\varphi} }{V_{(0)}} \frac{V_{(0)
*}}{V _ {,\varphi *}}\right)^2\right) \frac{V_{(0)}^2W_{(0)}^2
V_{(0)*}W_{(0)*} }{6(V_{, \varphi}^2W^2 +V^2W_{,\chi}^2)k^3}, \\
P_\zeta(\mathbf{k}) &=& (8 \pi G )^3 \left(\sin^2 \Theta\left(
\frac{W_{,\chi}}{W_{(0)}}\frac{W_{(0)*}}{W _ {,\chi *}} \right)^2
+\cos^2 \Theta\left( \frac{V_{,\varphi} }{V_{(0)}}
\frac{V_{(0)*}}{V _ {,\varphi *}}\right)^2\right)
\frac{V_{(0)}^2W_{(0)}^2 V_{(0)*}W_{(0)*} }{6(V_{, \varphi}^2W^2
+V^2W_{,\chi}^2)k^3}.
\end{eqnarray}

The inequality (\ref{largefnls}) can be rewritten now as
\begin{equation}
\label{neq1} \left|\left(
\frac{W_{,\chi}}{W_{(0)}}\frac{W_{(0)*}}{W _ {,\chi *}} \right)^2
-\left( \frac{V_{,\varphi} }{V_{(0)}} \frac{V_{(0)*}}{V _
{,\varphi *}}\right)^2\right| \gg |\tan\Theta|\left(
\frac{W_{,\chi}}{W_{(0)}} \frac{W_{(0)*}}{W _ {,\chi *}} \right)^2
+|\cot\Theta|\left( \frac{V_{,\varphi} }{V_{(0)}}
\frac{V_{(0)*}}{V _ {,\varphi *}}\right)^2.
\end{equation}
This condition can be satisfied for some inflationary models
\cite{BCH}.

\subsection{Sum potential, $U=V(\varphi) + W(\chi)$.}

In terms of initial fields perturbations, the quantities
$\delta\tilde{\sigma}_{(1)}$ and $\delta \tilde{s} _{(1)}$ are
\begin{eqnarray}
\delta \tilde{s}_{(1)}  &=& 8\pi G
\frac{(V_{(0)}+W_{(0)})V_{,\varphi} W_{,\chi}}{W_{,\chi}^2
+V_{,\varphi}^2 } \left(\frac{\delta \varphi_{(1)*}}{V_{,\varphi
*}} -\frac{\delta\chi_{(1)*}} {W_{,\chi *}}\right), \\
\delta \tilde{\sigma}_{(1)} &=& 8\pi G \frac{ W_{,\chi}^2V_{(0)} -
V_{,\varphi}^2W_{(0)}}{W_{,\chi}^2 +V_{,\varphi}^2 }
\left(\frac{\delta \varphi_{(1)*}}{V_{,\varphi *}}
-\frac{\delta\chi_{(1)*}} {W_{,\chi *}}\right) - 8\pi G \left(
\frac{V_{(0)*}\delta \varphi_{(1)*}}{V_{,\varphi *}}
+\frac{W_{(0)*} \delta \chi_{(1)*}} {W_{,\chi *}}\right).
\end{eqnarray}

The equation (\ref{zeta2sum}) yields
\begin{eqnarray}
\zeta_{(2)} &=&  \left[ \frac{V_{,\varphi*}^2 \!+\!
W_{,\chi*}^2}{8\pi G U_{(0)*}^2}\!-\!\frac{ V_{, \varphi}
^2(W_{(0)}\!+\!V_{(0)*}) \!-\!W_{,\chi}^2(V_{(0)} \!-\!V_{(0)*})
}{8\pi G U_{(0)*}^2 \left(W_{,\chi}^2 \!+\!V_{,\varphi}^2
\right)}V_{,\varphi \varphi*} \!-\! \frac{ W_{,\chi}^2(V_{(0)}
\!+\!W_{(0)*}) \!-\!V_{,\varphi} ^2(W_{(0)} \!-\!W_{(0)*})  }{8\pi
G U_{(0)*}^2\left(W_{,\chi}^2
\!+\!V_{,\varphi}^2\right)}W_{,\chi\chi *}
\right] \zeta_{(1)}^2 \nonumber \\
&-& 2\left[ \frac{V_{,\varphi \varphi *} \left(V_{, \varphi}
^2(W_{(0)}\!+\!V_{(0)*}) \!-\! W_{,\chi} ^2(V_{(0)}\!-\!
V_{(0)*})\right) \!-\! V_{,\varphi *}^2 \left(V_{,\varphi}^2 \!+\!
W_{,\chi}^2\right)}{8\pi G U_{(0)*}^2 U_{(0)}^2V_{,\varphi}^2
W_{,\chi}^2}\left( W_{(0)*} \!+\!\frac{ W_{,\chi}^2V_{(0)} \!-\!
V_{,\varphi}^2W_{(0)}}{\left(W_{,\chi}^2 \!+\!
V_{,\varphi}^2 \right)}\right) \right. \nonumber \\
&-& \left. \frac{ W_{,\chi\chi *} \left(  W_{,\chi}^2(V_{(0)}
\!+\! W_{(0)*}) \!-\!V_{,\varphi} ^2(W_{(0)}\!-\!W_{(0)*})\right)
\!-\! W_{,\chi *}^2 \left(V_{,\varphi}^2 \!+\!
W_{,\chi}^2\right)}{8\pi G U_{(0)*}^2 U_{(0)}^2 V_{,\varphi}^2
W_{,\chi}^2} \left(V_{(0)*} \!-\! \frac{ W_{,\chi}^2V_{(0)} \!-\!
V_{,\varphi}^2W_{(0)}}{\left( W_{,\chi}^2 \!+\! V_{,\varphi}^2
\right)} \right)\right]\zeta_{(1)}
\delta \tilde{s}_{(1)} \nonumber \\
&-& \left[ \frac{ V_{, \varphi} ^2(W_{(0)}+V_{(0)*}) -W_{,\chi}
^2(V_{(0)}-V_{(0)*}) }{8\pi G U_{(0)*}^2 U_{(0)}^2V_{,\varphi}^2
W_{,\chi}^2\left( W_{,\chi}^2 +V_{,\varphi }^2 \right)}V_{,\varphi
\varphi *}\left( W_{(0)*}\left( W_{,\chi}^2 +V_{,\varphi}^2
\right) + W_{,\chi}^2V_{(0)} - V_{,\varphi}^2
W_{(0)} \right)^2\right. \nonumber \\
&+& \frac{ W_{,\chi}^2(V_{(0)} + W_{(0)*}) -V_{, \varphi}
^2(W_{(0)}-W_{(0)*}) }{8\pi G U_{(0)*}^2U_{(0)}^ 2V_{,\varphi}^2
W_{,\chi}^2 \left( W_{,\chi}^2 +V_{,\varphi}^2 \right)}
W_{,\chi\chi *}\left(V_{(0)*} \left( W_{,\chi}^2 +V_{,\varphi}^2
\right)-W_{,\chi}^2V_{(0)} + V_{,\varphi}^2W_{(0)}\right)^2
\nonumber \\
&-&  \frac{ \left( W_{(0)*}\left( W_{,\chi}^2 +V_{,\varphi}^2
\right) + W_{,\chi}^2V_{(0)} - V_{,\varphi}^2W_{(0)}
\right)^2}{8\pi G U_{(0)*}^2 U_{(0)}^2V_{,\varphi}^2 W_{,\chi}^2}
V_{,\varphi *}^2 -\frac{  \left(V_{(0)*} \left( W_{,\chi}^2
+V_{,\varphi}^2 \right)-W_{,\chi}^2V_{(0)} + V_{,\varphi}^2
W_{(0)}\right)^2}{8\pi G U_{(0)*}^2U_{(0)}^ 2V_{,\varphi}^2
W_{,\chi}^2 } W_{,\chi *}^2\nonumber \\
&+& \left.\frac{V_{,\varphi}^2+W_{,\chi}^2}{8\pi G U_{(0)}^2} -2
\frac{ V_{,\varphi\varphi}W_{,\chi}^2 +W_{,\chi\chi} V_{,\varphi}
^2}{ 8\pi G U_{(0)}(V_{,\varphi}^2+W_{,\chi}^2)} \right]\delta
\tilde{s}_{(1)}^2 -\delta\tilde{\sigma}_{(2)}.
\end{eqnarray}

This equation leads to the same large $ f_ {NL} $ condition
(\ref{largefnl}), as in the case of product potentials. For the
case of sum potentials, we have
\begin{eqnarray}
K(\mathbf{k}) &=& \frac{(8 \pi G )^3 U_{(0)} U_{(0)*}V_{,\varphi}
W_{,\chi} }{6kV_{,\varphi *}^2W_{,\chi *}^2}\frac{V_{,\varphi
*}^2+W_{,\chi *}^2}{ V_{, \varphi}^2 +W_{,\chi}^2} \left( \frac{
W_{,\chi}^2V_{(0)} - V_{,\varphi}^2W_{(0)}}{ W_{,\chi}^2
+V_{,\varphi}^2 }- \frac{V_{(0)*}W_{,\chi *}^2 - W_{(0)*}
V_{,\varphi
*}^2}{V_{,\varphi *}^2+W_{,\chi *}^2} \right), \\
P_\zeta(\mathbf{k}) &=& \frac{(8\pi G)^3U_{(0)*}}{6kV_{,\varphi
*}^2W_{,\chi *}^2(W_{,\chi}^2 +V_{,\varphi}^2 )} \Big[ \left(
W_{,\chi}^2V_{(0)} - V_{,\varphi}^2W_{(0)}\right)^2 \frac{W_{,\chi
*}^2 + V_{,\varphi *}^2}{W_{,\chi}^2 +V_{,\varphi}^2} \nonumber \\
&-& 2\left( W_{,\chi}^2V_{(0)} - V_{,\varphi}^2 W_{(0)} \right)
\left(V_{(0) *}W_{,\chi *}^2 -W_{(0)*} V_{, \varphi *}^2\right) +
\left(V_{(0)*}^2 W_{,\chi *}^2+W_{(0)*} ^2V_{,\varphi
*}^2\right)(W_{,\chi}^2 +V_{,\varphi}^2 ) \Big] .
\end{eqnarray}

The inequality (\ref{largefnls}) gives
\begin{eqnarray}
\left| \frac{ W_{,\chi}^2V_{(0)} - V_{,\varphi}^2W_{(0)}}{
W_{,\chi}^2 + V_{,\varphi}^2 } - \frac{V_*W_{,\chi *}^2 - W_*
V_{,\varphi *}^2 }{V_{,\varphi *}^2+W_{,\chi *}^2}  \right|~~\gg
~~\left| \frac{ W_{,\chi}^2 +V_{,\varphi}^2}{U_{(0)} V_{,\varphi}
W_{,\chi}} \right|\left| \left( \frac{ W_{,\chi}^2V_{(0)} -
V_{,\varphi}^2W_{(0)}}{ W_{,\chi}^2 +V_{,\varphi}^2 }\right)^2
\frac{W_{,\chi *}^2 + V_{,\varphi *}^2}{V_{,\varphi *}^2 +
W_{,\chi *}^2}\right. \nonumber \\
\left. - 2\frac{ W_{,\chi}^2V_{(0)} - V_{,\varphi}^2 W_{(0)}
}{V_{,\varphi }^2 +W_{,\chi }^2 } \frac{V_{(0) *}W_{,\chi *}^2
-W_{(0)*} V_{, \varphi *}^2}{V_{,\varphi *}^2+W_{,\chi *}^2 } +
\frac{V_{(0)*}^2 W_{,\chi *}^2+W_{(0)*} ^2V_{,\varphi
*}^2}{V_{,\varphi *}^2+W_{,\chi *}^2 } \right| .
\end{eqnarray}
Some examples of slow-roll models satisfying this condition are
presented in ref. \cite{BCH}.

In general, the coefficients $C_{\tilde{s} \tilde{ \sigma}}$,
$C_{\tilde{s}\tilde{s}}$ do not have to be small. For instance, if
the field $\chi$ is subdominant during slow roll and
\begin{equation}
W_{(0)} \leq W_{(0)*}, ~~~~~ \left|\frac{ W_{,\chi} }{
V_{,\varphi} }\right| \ll 1 , ~~~~~\frac{ W_{,\chi}^2 }{
V_{,\varphi}^2 } V_{(0)}\ll W_{(0)*},
\end{equation}
then
\begin{eqnarray}
C_{\tilde{\sigma}\tilde{s}} &=& \frac{V_{, \varphi} }{W_{,\chi}
}\frac{ W_{(0)*}}{ U_{(0)} } \left(\frac{V_{,\varphi \varphi *}
}{8\pi G U_{(0)*}}  - \frac{ V_{,\varphi *}^2 }{8\pi G U_{(0)*}^2
}  - \frac{ W_{,\chi\chi *} }{8\pi G U_{(0)*}} \right) , \\
C_{\tilde{s}\tilde{s}} &=& -\frac{V_{, \varphi} ^2}{W_{,\chi}^2 }
\frac{W_{(0)*} U_{(0)*}}{ U_{(0)}^2 } \frac{W_{,\chi\chi *}}{8\pi
G U_{(0)*}  } ~.
\end{eqnarray}

The coefficients $C_{\tilde{s} \tilde{ \sigma}}$,
$C_{\tilde{s}\tilde{s}}$ can be valuable due the enhancing factor
$ V_{, \varphi} / W_{,\chi}$. In this regime, the curvature
perturbation $ \zeta^{(2)} $ remains constant and entropy
perturbation may be small, $ P_{\delta s} (\mathbf {k}) \ll P_
{\delta \sigma} (\mathbf{k}) $. An example of such solution in the
"adiabatic limit" was presented in ref. \cite{11062153}. Although
the entropy perturbation $\delta \tilde{s}_{(1)}$ is decaying, it
can not be set equal to zero, since then the parameter $f_{NL}$
gets a wrong small value.

\section{Conclusion.}
\label{concl}

We presented a derivation of general expression for the second
order curvature perturbation  $ \zeta^{(2)} $ in the form of a
functional over a background solution. For two special cases, the
explicit expressions was obtained that reproduce known results. It
is shown that, in contrast to the linear perturbation theory, the
curvature perturbation depend on both the adiabatic and
nonadiabatic perturbations. Moreover, a significant level of
non-Gaussianity can be generated  during the large scale evolution
only if nonadiabatic perturbations are non-negligible at the end
of inflation. Hence, to compare inflationary theory predictions at
large $f_{NL}$ with observational data, one need to investigate
the evolution of the large scale curvature perturbation on
post-inflationary stages.

\acknowledgments

The author is grateful to Joseph Elliston and David Mulryne for
useful discussions and comments.

{}

\end{document}